\documentclass[12pt]{iopart}
\usepackage[utf8]{inputenc}
\usepackage{graphicx}
\usepackage{lmodern}
\usepackage{color}
\usepackage{booktabs}
\usepackage{calc}
\usepackage{hyperref}
\usepackage{placeins}
\usepackage{subfig}
\usepackage{array}
\usepackage{times}
\usepackage{cancel}
\usepackage{harvard}

\newcommand{\jh}[1]{{\color{black}#1}} 
\newcommand{\lt}[1]{{\color{black}#1}} 
\newcommand{\jfd}[1]{{\color{black}#1}} 

\newcommand{\ltn}[1]{{\color{black}#1}} 
\newcommand{\JH}[1]{{\color{black}#1}} 

\newcommand{\ltnn}[1]{{\color{black}#1}} 

\newcommand{\sout}[1]{}

\begin{document}

\title[Modeler's guide to resilience]{\jh{A modeler's guide to studying the resilience of 
social-technical-environmental systems}}

\author{%
Lea A. Tamberg%
$^{1,2}$,
Jobst Heitzig%
$^3$ and 
Jonathan F. Donges%
$^{2,4}$
}

\address{$^1$
Institute of Environmental Systems Research, University of Osnabr\"uck, Barbarastraße 12, 49076 Osnabr\"uck, Germany}

\address{$^2$
FutureLab Earth Resilience in the Anthropocene, Earth System Analysis, Potsdam Institute for Climate Impact Research, Member of the Leibniz Association, Telegrafenberg A31, 14473 Potsdam, Germany}

\address{$^3$
FutureLab on Game Theory and Networks of Interacting Agents, Complexity Science, Potsdam Institute for Climate Impact Research, Member of the Leibniz Association, Telegrafenberg A31, 14473 Potsdam, Germany}

\address{$^4$
Stockholm Resilience Centre, Stockholm University, Kr\"aftriket 2B, 114 19 Stockholm, Sweden}

\ead{lea.tamberg@yahoo.de, donges@pik-potsdam.de}

\vspace{10pt}
\begin{indented}
\item[]\today
\end{indented}

\begin{abstract}
The term `resilience' is increasingly being used in \jfd{social-technical-environmental} systems sciences and particularly also in the Earth system sciences. However, \jfd{the diversity of resilience concepts} and a certain (sometimes intended) openness of proposed definitions can lead to misunderstandings and may impede their application to complex systems modelling. We propose a guideline that aims to ease communication as well as to support systematic development of research questions and models in the context of resilience. It can be applied independently of the modelling framework or underlying 
theory of choice. At the heart of this guideline is a checklist consisting of four questions to be answered: (i) Resilience of what? (ii) Resilience regarding what? (iii) Resilience against what? (iv) Resilience how? We refer to the answers to these resilience questions as the ``system'', the ``sustainant'', the ``adverse influence'', and the ``response options''. The term `sustainant' is a neologism describing the feature of the system (state, structure, function, pathway, \dots) that should be maintained (or restored quickly enough) in order to call the system resilient.

The use of this proposed guideline in the field of Earth system resilience is demonstrated for the application example of a potential climate tipping element: the Amazon rainforest.
The example illustrates the diversity of possible answers to the checklist's questions as well as their benefits in structuring the modelling process. The guideline supports the modeller in communicating precisely what is actually meant by `resilience' in a specific context. This combination of freedom and precision could help to advance the resilience discourse by building a bridge between those demanding unambiguous definitions and those stressing the benefits of generality and flexibility of the resilience concept.

\end{abstract}
\submitto{\ERL}

\section{Introduction}

The \jh{concept of `resilience'} \jfd{, broadly describing the capacity of a system to absorb or recover from perturbations,} is well-established in disciplines like psychology \cite{masten2002resilience,yates2004fostering}, ecology \cite{holling1973resilience,gunderson2012foundations}, materials science \cite{campbell2008elements}, and engineering sciences \cite{woods2006engineering,woods2015four,yu2020toward}. Resilience thinking has become an important guiding framework in research on social-ecological systems \cite{folke2010resilience,folke2016social} and in complex \jfd{(adaptive)} systems science more generally \cite{fraccascia2018resilience,lade2019comment}. 
Also, it \jfd{is increasingly being applied in Earth system science \cite{rockstrom2009safe,folke2010resilience,steffen2015planetary,gleeson2020illuminating,rockstrom2021identifying}} with a particular emphasis on Earth systems that have the potential to display critical thresholds and tipping points \cite{lenton2008tipping,dakos2015resilience,wunderling2020basin}.

In this interdisciplinary context, a vast number of definitions, concepts and related terms has been proposed \jh{to capture various aspects of resilience}. For instance, the relation between \jh{terms such as `stability', `adaptability' and `transformability'} is intensely debated in the context of ``resilience thinking'' \cite{folke2010resilience,cote2012resilience,walker2012resilience,curtin2014foundations,donges2017math,lade2017modelling}.  

Part of the theoretical discussion is whether a narrow or a broad definition of the term `resilience' \jfd{is preferable}. In ecology, a \jfd{narrower scope appears to be emphasized} \cite{holling1973resilience,brand2007focusing,kefi2019advancing,Hodgson2015}, often seeing `stability' as the more general and `resilience' as the more narrow term \cite{vanmeerbeek2021}. Other authors, especially from the domain of social-ecological systems, explicitly advocate for a broader understanding of the term \cite{folke2010resilience,walker2012resilience,anderies2006fifteen}, \lt{ valuing its role as a boundary object or bridging concept between different academic and non-academic fields \cite{baggio2015boundary,Turner2010a}.}

At the same time, resilience concepts may be hard to apply, operationalize or quantify when it comes to the analysis and modelling of specific \jfd{real-world complex systems, particularly beyond well-quantified scientific fields such as physics or materials science}. This has several reasons including the mentioned \jfd{openness of definitions}, lacking formalization, and missing estimation methods  \cite{strunz2012conceptual,brand2007focusing,Hodgson2015}. Also, experience shows that research questions based on abstract theoretical concepts \jfd{often} cannot easily be answered with \jfd{pre-existing models} that were not specifically developed for this purpose. Often, central aspects of resilience theory are simply not represented in the model. For instance, a model in which the possibility of structural changes is not included does not fit to a research question addressing the adaptation or transformation capacity of a system.

\jh{These considerations imply at least two complementary challenges for the modeller.}
First, the large variety of previously proposed definitions and theoretical frameworks in different fields makes it necessary to communicate very precisely what is actually meant by `resilience' in a particular \jh{study} to avoid confusion. 
Second, research questions and the model(s) to answer them should be developed simultaneously in the same structured process in order to ensure their compatibility.

\jfd{In this paper, we argue that these two concerns---precise communication and compatible modelling---can be addressed together \jh{in the modelling process. For this, we propose a guideline for an iterative approach to modelling systems\footnote{These systems could be of any kind, from purely physical to social-technical-ecological. The resilience perspective is particularly promising for the latter; However, the guideline should be applicable to all domains of systems modelling.} for studying their resilience.}} \jfd{Our approach is based on a checklist that helps narrowing down on the precise form of resilience to be studied, the respective research question to be answered, and how a model should be designed in order to address that question.} \jh{We exemplify our approach at the hand of three illustrative example applications.}

\jh{In view of the ongoing theoretical debate, we take a neutral position regarding the definition of debated terms such as `resilience', `adaptability', `stability', `transformability', etc.}

\jh{This way, the modeller does not need to decide on one particular theoretical framework upfront but may do so only later when interpreting their results in one or other such framework.}

The style of the questions of the checklist is inspired by \cite{carpenter2001metaphor}, who demand to answer the question ``Resilience of What to What?'' \jfd{Our} guideline extends the required precision by differentiating this question \jh{considerably.} 
Also, we put less conceptual restrictions on the possible answers to the questions. \ltn{Related question lists have also been proposed for ``politics of urban resilience'' \cite{meerow2019urban} and for ``stability'' in ecology \cite{grimm1997babel}}\JH{, which are either conceptually similar or rather complementary to our checklist}.

\jh{The paper is structured as follows: In Sect.\ \ref{sec:method}, we present a proposed methodology for resilience modeling. By way of results, we then examplarily apply this methodology to the Amazon rainforest in Sect.\ \ref{sec:amazon}, which is complemented by two more application examples in the Appendix. The discussion in Sect.\ \ref{sec:discussion} concludes the paper.}

\section{A methodology for resilience modeling \jh{and research question refinement}}
\label{sec:method}

\begin{figure}[ht!]
\centering
\includegraphics[width=0.8\textwidth]{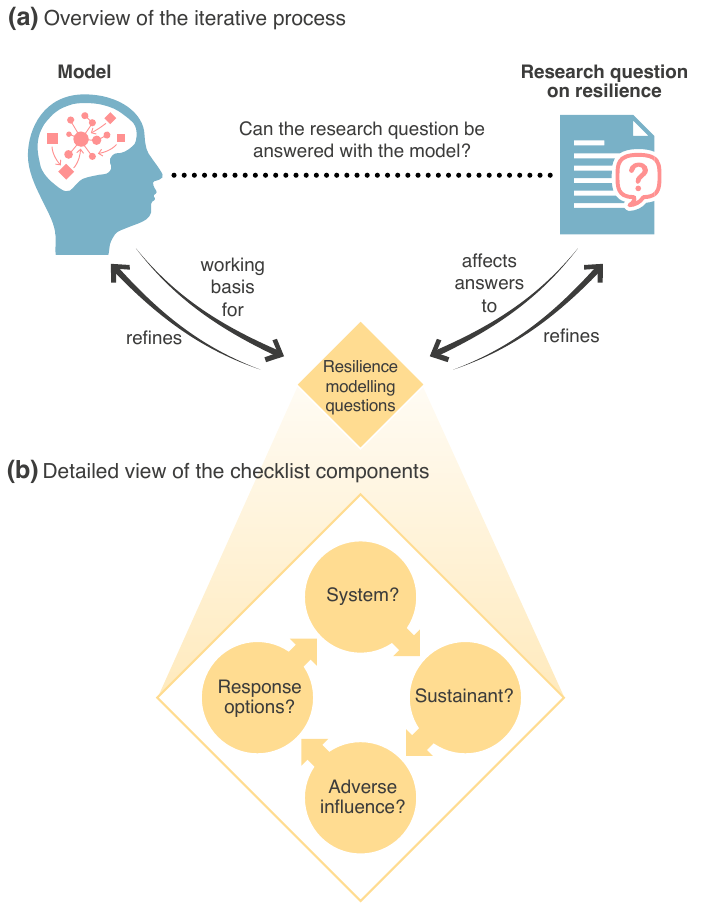}
\caption{\jfd{Illustration of the iterative resilience modelling process suggested when using the proposed guideline (a). Starting from a basic (mental) model and a broad research interest, the checklist of four resilience questions (b) can help to refine both until they are internally consistent and sufficiently concrete for simulation modelling.}\label{fig:guideline}}
\end{figure}

\jfd{In this section, we introduce and justify our proposed framework for \jh{developing simulation models of large social-technical-environmental system, such as} complex Earth systems\jh{, to be used to answer different kinds of research questions relating to the system's resilience, such as quantifying it, assessing its evolution over time, determining factors that influence it, or identifying ways to improve it}.}
\jh{The approach we put forward can be seen as an iterative process, summarized in figure \ref{fig:guideline}. }
\jh{We start from the assumption that in any particular research study, at least a basic, possibly only mental, initial ``model'' of the system exists, as well as an initial, possibly only broadly defined, research interest.} They serve as a working basis for the further process.
While working through \jh{a checklist of guiding questions,} more precise characterizations for an improved model and a more specific formulation of the research question will arise. Whenever the answer to one guiding question is modified, it may become
necessary to reconsider certain other guiding questions iteratively until the system model and research question are consistent. To arrive at this point, both a solid domain expertise and sufficient methodological knowledge about the ingredients of our guideline are needed, but no commitment to a specific theoretical resilience framework.

In the following, the different questions of the checklist are presented and explained with the help of examples. Each of these questions includes a set of subquestions helping to be as precise as possible. The general idea is to specify four different aspects of resilience in the context of a specific system: {\em Resilience of what, resilience regarding what, resilience against what and resilience how?}

\subsection{Resilience of what: What is the system?}

What are the system \textit{boundaries} and how sharp are they? What are the system’s \textit{parts} and their \textit{interactions}
that appear relevant for answering the research question? 
With what aspects of its \textit{environment} does the system interact, through which kind of {\em interfaces?} Is there \textit{agency}\footnote{Agency is originally a concept from sociology \cite{barker2002making}, meaning the capacity of individuals to act independently and to make their own free choices. It is increasingly used in the context of resilience in social-ecological systems \cite{armitage2012interplay,larsen2011governing,westley2013theory,otto2020human}.}
in the system, meaning that parts of the system may exhibit targeted, intentional action?

\lt{Following the usual linguistic convention, the question ``Resilience \textit{of} what?'' refers to the whole system of interest, not a specific system state as in \cite{carpenter2001metaphor}. The latter is covered in the next question of the checklist.} Working out the mentioned aspects of the system is of course not a step only taken in the resilience context but in systems modelling in general \cite{bossel2007systems,voinov2010systems}. \ltnn{Particularly, the choice of boundaries is a typical challenge and often needs to be modified with a changing research interest (see the application examples in section \ref{sec:results} and the appendix for typical choices}). The following questions are more focused on the modelling of resilience itself and build on the results of the first one, often making it necessary to reconsider those in further iterations.

\subsection{Resilience regarding what: What is the “sustainant”? }

Which feature or property of the system is \textit{supposed to be sustained} \jh{or maintained} in order to call the system resilient? Its state or structure, its pathway? Some long-term equilibrium? Its function, purpose, \jh{or utility for some stakeholder}? Some quantitative or qualitative aspect of the system? What this ``sustainant'' is is no objective feature of a system but is normatively chosen by the observer from their perspective, which should be clearly communicated. Especially what the ``function'' or ``purpose'' of a system is can be seen differently from different perspectives \cite{cutter2016resilience,meerow2019urban}.

For instance, for different observers, the function of a forest could (among others) be to produce wood, to enhance biodiversity, to provide a habitat, to serve for recreation, or to be beautiful. The model analysing the resilience of the forest regarding wood production would differ substantially from a model with biodiversity as the sustainant. 

\lt{Note that the neologism `sustainant' is conceptually distinct from the broader concept of `sustainability' \cite{anderies2013aligning}. Instead, it is intended to cover different ideas from the literature about which system property the resilience of a system is related to.  For instance, \cite{carpenter2001metaphor} demand the specification of a certain system state as an answer to the question ``Resilience \textit{of} what?'' (while their ``\textit{to} what'' refers to our ``\textit{against} what''). In contrast, \cite{folke2010resilience} define that a system is resilient if it essentially retains ``the same function, structure and feedbacks''. The term `sustainant' is not restricted to one of these (or other) perspectives but leaves it to the modeller to clearly specify which system property is of interest. 

Considering the normativity of the sustainant, it becomes clear that its choice can be subject to power relations, inequality, and competing interests. Many authors therefore demand to consider the question ``Resilience for whom?'' to account for these aspects \cite{cretney2014resilience,cutter2016resilience,meerow2019urban}. This question is located on a meta level above that of the checklist. It can help to both choose the sustainant and to criticise this choice, for example from an inequality perspective.}

Part of the task of selecting the sustainant is to ask whether there are any kinds of {\em threshold} values for certain indicators that shall \jh{either not be exceeded ever or may only be exceeded temporarily.}
For example, when modelling the development of the oxygen concentration in an aquarium and choosing this system \jh{property} 
as the sustainant, its restoration after a drop may be irrelevant if it was zero in between so that all fish have died. One could argue that in this example, a better sustainant would be the fish being alive. However, this is a question of model boundaries. If the fish is not explicitly modelled but is only described as a consumption factor in the water-oxygen system, the potential interest of the modeller in the fish staying alive leads to the definition of an acceptable range for the oxygen level (possibly even different ranges for different species). If the modeller does not care about fish survival but the general capacity of the system to restore oxygen level, the sustainant can be defined without any threshold values.

Such thresholds depend, as the choice of the sustainant itself, on the perspective and interests of the observer who has to define an \textit{acceptable range} for the sustainant. Correspondingly, an \textit{acceptable recovery time} should be defined.

For instance, a fish stock may recover 50 years after a collapse; however, this is not relevant for someone aiming to evaluate the risks of investments into the fishery industry that is dependant on this resource on much shorter time scales. Again, the boundary choice (here only to model the fish population and describe the fishery as an external factor) leads to the population size being the chosen sustainant, specified by an acceptable recovery time motivated by the concern for the fishery industry.

As another example, consider a social network. A possible sustainant could be that every individual has at least one connection  
to another individual. This sustainant would be a property of the system's structure. An appropriate recovery time could for instance consider how long an individual can endure social isolation without developing mental illness. Of course, this could also differ from individual to individual.

\subsection{Resilience against what: What is the adverse influence?}

What is the concrete influence \textit{affecting the sustainant} that shall be considered for this specific resilience analysis? Is it an abrupt but temporary disturbance (pulse), a shock, a constant pressure, noisy fluctuations, a perturbation, an abrupt but permanent shift in some feature, 
or a slow change? Does it originate in the system or in its environment? Does it affect the structure, a parameter, or the state of parts of the system? 

In some cases, the influence that is supposed to be studied does not have a direct effect on modelled aspects of the system, but through an intermediate linked to the boundary interface of the system, making it important to be precise about the actually modelled influence. 
Note that the term `adverse' in `adverse influence' is not necessarily meant as something undesirable. It only reflects that this influence effects the system in a way that weakens the sustainant. If the observer views the sustainant as undesirable, the adverse influence on it can be seen a positive process \cite{donges2017math,dornelles2020towards}.

\medskip
Remark to the 2nd and 3rd question of the checklist (sustainant and adverse influence): Of course, a sustainant could be composed of several aspects and resilience could be required against different (internal or external) influences at the same time. Often, such a “multi-resilience” is of higher interest than a “single\jh{-aspect}-resilience” regarding only one parameter. However, the required analysis is more complex, not only because more aspects have to be modelled and studied but also because possible interdependencies between different sustainants and/or influences must be considered \jh{and the design of a suitable aggregate quantitative indicator may be more difficult and risk to appear arbitrary}.\footnote{Folke et al. distinguish between \textit{specified resilience} (``resilience of some particular part of a system, related to a particular control variable, to one or more identified kinds of shocks'') and \textit{general resilience} (``resilience of any and all parts of a system to all kinds of shocks, including novel ones'') and argue to concentrate on the latter one in order to cope with uncertainty and trade-offs \cite{folke2010resilience}. However, in a modelling context, specificity is crucial. It is inherent to the modelling process that decisions on what to represent in the model and what not have to be taken. Therefore, our guideline asks to \textit{specify} which sustainants and influences are considered. By choosing a multi-sustainant and a multi-influence, the risk of overlooked trade-offs can at least be reduced.}

For instance, one may be interested in the resilience of the climate system against a rise of CO$_2$ concentration in the atmosphere (= single-influence) regarding the global mean temperature (= single-sustainant) or regarding the ensemble of temperature, precipitation, and wind maxima over the course of the year in each region (= multi-sustainant).

Another example of interest may be the resilience
of a society against increasing abundance of misinformation and the shock of a pandemic (multi-influence) regarding trust in the government (single sustainant) \cite{bak2021stewardship}.

\begin{figure}[h!]
    \centering
    \includegraphics[width=\textwidth]{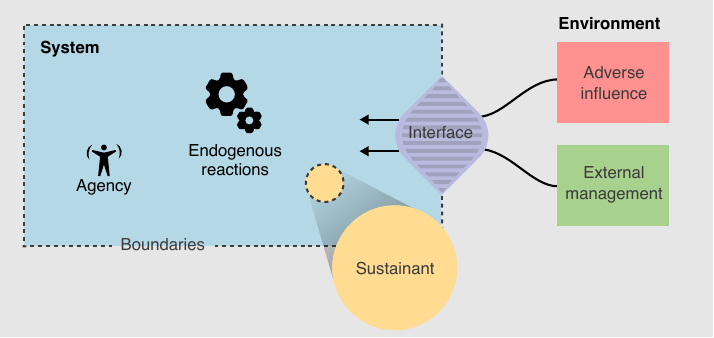}
    \caption{Relations between \jfd{central} terms such as `sustainant' and `adverse influence' as used in our \jfd{resilience modeling} guideline}
    \label{fig:terms}
\end{figure}

\subsection{Resilience how: What are the response options?}

At which levels can or does a system react to \jh{adverse influences}? Which types of reactions can be observed? (One type could be the inherent stability behaviour of a system, bringing it back to a state compatible with the sustainant (for instance in a fish population with logistic growth). Another response could be a change of structure or rules, e.g., the creation of new links in a communication network or the switch to another set of rules in a transport system.) What is the range of possible reactions? Which reactions are endogenous as a consequence of the system’s structure and rules? Which response options require external management or agency? In figure \ref{fig:terms}, some important terms used in the checklist are presented graphically.

\subsection{Refining the research question on resilience}

With the help of the checklist and the \jh{notion of `sustainant',} the research question on system resilience can be specified more precisely. \jh{Some possible types of such research questions are:} 

\begin{itemize}
    \item {\em Is the system \jh{``in general''} resilient regarding the sustainant?} This might be a rather qualitative question: Is the sustainant easily affected and does it recover in an acceptable time range? 
    \item {\em How much influence can the system bear without a change of the sustainant or with a recovery on a relevant time scale?} This corresponds to a quantified measure of the specific form of resilience analysed,
    \jh{e.g., using metrics such as ``basin stability'' or ``survivability'' or variants thereof \cite{menck2013basin,Mitra2015,Hellmann2016,VanKan2016}}. 
    \item {\em How can the system be designed to be more resilient through its structure and \jh{internal dynamical} rules?} Often, this is a question for general rules about how to build or fix certain kinds of systems so that they show the desired resilience \cite{biggs2015principles}. To answer this question, the system model(s) must have a certain level of genericity \jh{that allows for the comparison of different structural or dynamical changes}.
    \item {\em What are resilience-promoting management 
    \jh{strategies?}} In this case, the model should reflect the possibility to change structure, states, parameters and rules/cause-effect relationships easily.
\end{itemize}

\subsection{Choosing appropriate modelling techniques}

Answering the above guiding questions 
does not produce a complete model but a collection of requirements that should be met by a more technical description. For this, our guideline does not specify a single approach. In general, any mathematical or simulation technique from differential equations over agent-based modelling to game-theoretical modelling 
may be used.
Of course, the description resulting from answering the guiding questions will influence this choice. For instance, if an important feature of a system is the social structure connecting people, choosing a network model appears natural. 

\section{Application example:
Amazon rainforest}
\label{sec:amazon}
\label{sec:results}

To show how the proposed guideline can be applied in modelling and communication of Earth system resilience, let us consider \jfd{the Amazon rainforest}: \jfd{its hyperdiverse ecosystem} and the human societies interacting with it.

We start with the broad research interest of whether climate impacts may cause a large-scale die-back of the forest via possible tipping dynamics \cite{lenton2008tipping,lovejoy2018amazon}.
To highlight the flexibility of the guideline, 
we formulate several related research questions and discuss respective modeling options. 
Each version is meant to be a potential result of one or more iterations of the above process.

\subsubsection*{1st version: Aggregate tree-cover reacting to overall aridity.}

From the broad research interest, one may derive the straightforward binary 
\textit{sustainant} that Amazonia remains a predominantly forested area. 
A possible quantitative indicator of this sustainant may be that the overall share of area that is covered by forest, $0 \le C \le 1$, is above 1/2.
A related \textit{system} model 
could have as its sole variable the rainforest cover, influenced by climate conditions. If water availability is seen as the limiting factor for vegetation growth which is most affected by climate,
the relevant \textit{adverse influence} is a potentially increasing aridity $A$, which could hence serve as the sole parameter of the model.
A rather simplistic example of such a model was given in \cite{menck2013basin} as
\begin{equation}
    dC/dt = - \delta\, C\: + \:1_{C > C_0(A)}\, \gamma \,C (1 - C),
\end{equation}
where $\delta$ is a decay rate due to respiration and degradation and $\gamma$ and $C$ are the parameters of a logistic growth happening when $C$ exceeds some minimal value $C_0(A)$ that depends on $A$ in a strictly monotonic fashion.
This model has a stable fixed point of $C^* = 1 - \delta / \gamma$  (largely forest) as long as $C_0(A) < C^*$, 
and another stable fixed point of $C = 0$ (pure savannah) as long as $C_0(A) > 0$.
If it is in the forest equilibrium, it is unaffected by changes in $A$ that keep $C_0(A) < C^*$, since its responses to those changes are not detailed in the model. 
Once $C_0(A)$ exceeds $C^*$, the model goes to the savannah equilibrium.

Here, the only modelled \textit{response option} of the system to the slow parameter change of $A$ is its inherent relaxation to equilibrium. Therefore, such a model may be used to answer \textit{research questions} such as: If there were no other influences than an increase in aridity, would the Amazon rainforest be resilient enough to survive predicted levels of global warming? How much can aridity increase without a die-back? \footnote{As the described model is extremely stylized and its quality mostly dependent on the quality of the function $ C_0(A)$, it should not be interpreted as a recommendation for answering the research questions but as an easily understandable example for the kind of models more suitable for this purpose.}

\subsubsection*{2nd version: Adding abrupt reductions of forest cover.}

In the above type of model, the only human impact on the system is increasing aridity. 
To get a more realistic picture of the risk humans put on the sustainant, one may consider other, often abrupt mechanisms on the system, such as droughts \cite{droughts}, deforestation \cite{deforestation}, fire \cite{fire}, storms \cite{storms} or soil poisoning by mining activities \cite{mining}. Formally, a \textit{multi-influence} could be defined, composed of increasing aridity and sudden forest loss. 
This would help formulating various \textit{research questions}, e.g. about the interplay between the two different influences: How does an increase in aridity shrink the basin of attraction of the forest state so that sudden forest losses get more likely to push the system to the savannah state (e.g. \cite{menck2013basin})? How much do these shocks have to be reduced by external management for any given level of climate change to avoid a collapse of the sustainant?

\subsubsection*{3rd version: Adaptation of species composition.}

Only considering the aggregate tree cover ignores crucial ecological adaptations of species composition \cite{species} and forest structure \cite{structure}.
Related \textit{research questions} are: How much do these additional \textit{response options} help sustaining or recovering the forest cover for any given level of the identified adverse influences \cite{sakschewski2016resilience}? At what rapidity of climate change will such adaptation get too slow to prevent collapse?
Studying these requires significant changes in the model, tracking stocks of different phyla, genera, species, etc., e.g., via coupled differential population equations or more sophisticated dynamic vegetation models \cite{shugart1984theory,kohler2003simulating,fischer2016lessons,botkin2007forecasting}. 

\subsubsection*{Further directions.}

As the Amazon rainforest interacts with both the climate system \cite{shukla1990amazon,cowling2008water} and with socio-economic systems \cite{muller2019can} via various, sometimes rather regional feedback mechanisms, examples of further {\em research questions} may be: How do the resiliences of the whole Amazon rainforest, its regional parts, and these coupled systems interact? How should rules ensuring a sufficiently low level of deforestation despite economic shocks be designed?

Addressing these research questions may require seeing the relevant {\em system} as including the moisture-recycling atmosphere \cite{zemp2017self} and/or certain land use systems, adding economic shocks as another adverse influence, regionally disaggregating both {\em adverse influences} and {\em response options,} and adding the corresponding spatial resolution to the model.

\section{Conclusion} 
\label{sec:discussion}

The guideline proposed above has been developed to support research on and communication about resilience in the field of complex systems at large \jh{and social-technical-environmental Earth systems in particular.} As the application example shows, the presented checklist is not a strict recipe but rather a guideline that can be used to structure the modelling process, ensuring that all important aspects are taken into account. Still more importantly, it helps to communicate clearly about the specific research question of interest and the meaning of the term `resilience' in the context of a specific system. \ltnn{All these aspects help the modeller to meet their responsibility for research quality, transparency, and replicability.}

Introducing the term `sustainant' allows to communicate about the system property of interest without predefining its nature (e.g., function, state, structure, \dots ). This enables the modeller to analyse resilience ``anchored in the situation in question'' as it has been demanded by \cite{grimm1997babel} for the term `stability'. \lt{Following the checklist, one can avoid getting lost in discussions about relationships between terms like `persistence', `adaptability', `resistance', etc. in the modelling process. Still, it is possible to reconnect a model developed following our checklist to different theoretical terms. Depending on the choice of sustainant, adverse influence, and response options, one can examine different definitions of resilience. For instance, according to \cite{Hodgson2015}, resilience is the resistance and/or recovery of a system's state, where resistance is the ``instantaneous impact of exogenous disturbance on system state'' and recovery ``captures the endogenous processes that pull the disturbed system back towards an equilibrium'' \cite{Hodgson2015}. This interpretation can be the result of a specific set of answers to our checklist's questions: the sustainant is the system's state, the adverse influence is some exogenous disturbance aiming at changing this state and the response options are both the capacity of the system to stay unchanged in state (or only change slightly) despite this disturbance (=resistance) and the capacity to develop back to the original state after the state has been perturbed (=recovery).
As a second example, consider the forms of resilience used in ``resilience thinking'': persistence, adaptability and transformability \cite{folke2010resilience}. A model with a specific system state as the sustainant can be used to study persistence, while adaptability requires the model to include the possibility to change the system's structure or rules. If one would like to examine the transformability of a system, it is obvious that the chosen sustainant would have to be a more abstract property, such as system function, since a system undergoing a general transformation of structure and feedbacks (e.g., according to \cite{folke2010resilience} a change of its stability landscape) can by definition not be resilient regarding for instance its pathway. These examples from resilience theory show how different answers to our checklist can be used to cover different abstraction levels of resilience, reaching from a simple ``bounce-back'' conception to the broader perspective of transformation.} \ltn{By this, our approach could help to build a bridge between two different views on the concept of resilience: as already mentioned in the introduction, some authors stress the importance of a clearly specified concept in order to facilitate formalization and measurement while others value resilience as a transdisciplinary bridging concept. Our guideline connects elements from both views. A modeller \textit{can} examine the system for its adaptability or resistance by choosing appropriate answers to the checklist's questions. However, in contrast to merely naming these terms, a conscientious application of the guideline will clarify and define their meaning in a specific context. Therefore, the proposed approach may help to contribute to the development of a unified theoretical framework on resilience of complex systems.}

The current version of our framework, as presented in this article, is still limited in several ways, which opens up respective avenues for future extensions and research. For example, we do not define any type of quantitative ``resilience metrics'' or study their consistency. However, our questions may help defining such metrics, which might take different mathematical forms to deal with these various dimensions, e.g., in a way similar to basin volume-based metrics proposed by \cite{menck2013basin,Mitra2015,Hellmann2016,VanKan2016} or the quantifiers following \cite{Hodgson2015}. One example of this is the work of \cite{bien2021resilience} in the context of power grids.
A more sophisticated type of novel resilience metric might be a real-valued function $f$ that maps a combination of four indicators---one for the current state $x$ of the system, one for the acceptable threshold $\theta$ of the sustainant, one for the strength $\sigma$ of potential adverse influences, and one for the allowable recovery time $T$---to the probability $p=f(x,\theta,\sigma,T)$ that the system will return to acceptable levels of the sustainant within the allowable time after suffering in the specified state an adverse influence of the given strength. This approach and other quantifiers would make an important contribution to the study of resilience in complex systems. 

Another limitation is that our avoidance to choose a specific theoretical resilience framework could be understood as a capitulation to the exhausting but important process of concept formation. Still, the application of a similar pragmatic approach such as ours to a series of concrete systems could eventually reveal general insights that might feed back into the more theoretical discussions about the concept of resilience.
For instance, conceptualizations along the dimensions of system, sustainant, adverse influence, and response options may be used in some kind of ``resilience study intercomparison project'' in the spirit of the inter-sectoral impact model intercomparison project (ISI--MIP) \cite{isimip} or similar activities. \jfd{Such an endeavour would be particularly helpful for choosing from the large group of existing Earth system models those that help best assessing the resilience of the Earth system regarding the sustainant of a `safe operating space' --- defined by planetary boundaries --- against the adverse influence given by human pressures such as greenhouse gas emissions and degradation of biosphere integrity.}

\ack

The idea for this work was developed during a ``Studienkolleg'' working group on ``The Earth as a complex system'' funded by the German National Academic Foundation. The work was carried out as part of the bachelor program on Applied Systems Science at Osnabr\"uck University together with the COPAN collaboration at the Potsdam Institute for Climate Impact Research. J.H.\ and J.F.D.\ are supported by the Leibniz Association (project DominoES), the European Research Council (ERC) under the European Union's Horizon 2020 Research and Innovation Programme (ERC grant agreement No.\ 743080 ERA). L.A.T. is supported by the German National Academic Foundation. 
We acknowledge further support by Deutsche Forschungsgemeinschaft (DFG) and the Open Access Publishing Fund of Osnabrück University. 
We thank Frank Hilker, J\"urgen Kurths, Steven J. Lade, Marc Wiedermann, Nico Wunderling, Vivian Z. Grudde, Bastian Grudde, Adrian Lison, Florian Schunck, and Sascha Haupt for valuable comments and discussions.

\section*{Bibliography}

\bibliographystyle{jphysicsB}
\bibliography{resilience_library.bib}

\clearpage
\markboth{Appendix}{}
\appendix

\section{Additional example: A fishery}
\label{sec:fishery}

As an additional application of much smaller extent than the amazon rainforest, let us study the paradigmatic and well-studied example from environmental economics of a fishery \cite{Perman2003}. 
The most elementary description of this system is a fish stock that is harvested by a fisher community (= basic mental model). Traditionally, one would like to analyze the ability of this system to maintain its harvested yield (= broad research interest).

\subsubsection*{1st version: Constant harvesting effort.}

\textit{What is the system?} A very simple way to model a fishery system is by a single differential equation describing the change of fish population size via logistic growth and harvesting. In this case, the system description only has one \jh{variable,} 
the fish stock. There are two interfaces to the system's environment. First, the \jh{fish} population, $x$, is influenced by ecosystem factors such as food supply, competition with other species and climatic conditions. All these aspects are aggregated in the two parameters, the intrinsic growth rate $r$ and the carrying capacity $K$. The second interface is the harvest of fish by human fishers, modeled as a subtractive harvesting term, e.g.\ a concave term controlled by an effort parameter $h$ \jh{and elasticity $\alpha<1$:}
\begin{equation}
    \frac{dx}{dt} =\, rx \left(1-\frac{x}{K}\right) - hx^\alpha.
    \label{eq:logistic}
\end{equation}

In this scope, the model does not reflect any agency---the harvest effort is a given for this choice of boundaries. 

\textit{What is the sustainant?} So far one can imagine different sustainants. An environmental organization could consider the system resilient if $x>x_{\min}$ for some threshold $x_{\min}$. In environmental economics, more importance is given to yield, $y = hx^\alpha$, which is what we choose here as well. 
One could define a minimal yield $y_{\min}$ and a maximal time $t_{\max}$ that $y$ may stay below $y_{\min}$ because fishers have limited financial reserves.

\textit{What is the adverse influence?} The sustainant can be challenged by an abrupt reduction of $x$ due to, e.g., a fish pest or an invasion by an external fishing fleet. Since a reduced $x$ means less yield, this affects the sustainant.

\textit{What are the response options?} The only response ``option'' of the system is the built-in basic dynamic stability that lets the stock converge to its unique stable equilibrium value \jh{$x^\ast$ from whatever initial condition $x(0)$. Depending on parameters and perturbation size, $x$ may recover fast enough to ensure that the drop in $y$} does not last too long.

\textit{Research question}: First of all, one can ask whether\jh{, given some value of $h$, equilibrium yield $y^\ast = h(x^\ast)^\alpha$ is above $y_{\min}$. If so, the model can be used to find out by how much $x$ may be reduced without exceeding the acceptable recovery time for $y$. With knowledge of the probability distribution of such reduction events, one may also calculate the risk and expected first occurrence time of a fishery breakdown depending on the value of $h$.}

\subsubsection*{2nd version: Adaptation of harvesting effort.}

In a more realistic model version, one could argue that fishers could adapt their \jh{effort $h$ to a changing $y$}. This extension of the \textit{response options} makes it necessary to modify the model. Instead of \jh{treating $h$ as an exogenous parameter,} we need a new component representing the harvest decisions of the fishing community. \jh{This could be done by specifying $h$ (or the change $dh/dt$) as a function $f$ of current and past yield, $y(\leq t)$.} 
The modified model now includes agency since it models the fishers' reaction to changing yield. A \textit{research question} could be: What is the optimal effort function $f$ that minimises the risk of collapse? 

\subsubsection*{3rd version: public good problem.}

The 2nd model would certainly lead to the insight that under certain conditions, $h$ has to be reduced temporarily or permanently to ensure a long term sustainable $y$. However, a fishing community is typically not a single entity but a heterogeneous group of fishers with individual interests. \jh{The model could thus reflect the resulting public good problem by including several agents $i$.} This extension results in changes on all levels of the checklist. In contrast to the first and second model, we now need to model \jh{agents' decisions on individual efforts $h_i$.} This could be done by assuming the same strategy for all fishers, e.g., individual short-term profit maximization. 
However, a large diversity of other approaches is possible, usually introducing more heterogeneity. \jh{Since our original \textit{sustainant}, overall yield, is indifferent to yield inequality between fishers, an alternative sustainant could be that for a specific percentage of fishers, individual yield $y_i$ does not collapse below some $y_{i,\min}$ for longer than some time $\Delta t_{i,\max}$. Or some welfare function $W(y_1,\dots,y_N)$ from welfare economics may be used to define a quantitative sustainant.} 
The \textit{adverse influence} would be the same as before but there would be several layers of \textit{response options}: \jh{The stock's inherent convergence to equilibrium, the effort change of the agents due to their existing strategies, the change of these individual strategies, e.g.\ due to social or individual learning, and the collective setting of rules by the community.} One may then have a suitable model to answer---amongst others---the following \textit{research question}: What rules should the community implement to maintain the sustainant?

~

\noindent
A {\it 4th version} might be necessary if fishers generate \jh{profit $y_ip-g(h_i)$ (the new sustainant, with $y_ip$ being the sales and $g(h_i)$ the costs given by some function $g(.)$) by selling their yield to a market (an additional model component) whose price $p$ may drop (adverse influence), to which they may react by jointly reducing their efforts $h_i$ to maintain a higher $p$ or by redistributing income through taxation (response options).}

\section{Additional example: An electricity transmission system}

Power grids have often been studied regarding various forms of stability, robustness, and performance (e.g., \cite{Menck2014,Plietzsch2015,Hellmann2016,Nitzbon2017c,Wienand2019}), many of which can naturally be seen as specific forms of resilience (e.g., \cite{anderies2013aligning}). Therefore, this kind of system is a 
good example to illustrate
the proposed framework,
\jh{even though it is only interacting with the environment but has no major environmental component itself}. A basic starting model description for a power grid could be that electricity producers and consumers are connected by power transmission lines. The broad research question is to analyze if the grid is easily disturbed by changing conditions.

\subsubsection*{1st \jh{version:} 
static consumption patterns.}

\lt{Differentiation between iteration of the checklist for one model version or several as here for didactic reasons}

\textit{What is the system?} The historically earliest and also most simple model of a power grid is a graph with edges representing high-voltage transmission lines and nodes representing transformers to lower voltage levels, aggregating consuming and producing subsystems not further defined. The interface to the environment of the system is the production/consumption of every node. Each transmission line has a certain capacity. It is assumed that production and consumption are always balanced. This reflects the fact that the model does not include mechanisms to match electricity offer and demand, such as a market. The model is non-dynamic, the electricity transport is calculated with static power flow equations basically representing Kirchhoff’s laws. This can be used as a base model to address the next questions of our framework. 

\textit{What is the sustainant?} From the perspective of a society maintaining a power grid, the function and sustainant of such a system can be seen as enabling all power transmission desired by producers and consumers. This is a binary sustainant: Either everyone's transmission demand is met or not.

\textit{What is the adverse influence?} In this model, the sustainant can be challenged by a new (but still balanced) production and consumption pattern that may lead to line overload. Since our base model treats production/consumption as part of the environment rather than as part of the system, this influence is seen as an external influence on the system at this point, affecting the system by the input or consumption status of nodes.

\textit{What are the response options?} Since the model does not include any agency, the only response option is that the power flow adjusts automatically to shift load from overloaded lines to others as a consequence of Kirchhoff's laws.

The model is suitable for answering the very specific \textit{research question}: “Can the grid transmit all desired production/consumption or not?” for a specific state, as well as deriving from that: “Which production/consumption patterns’ transmission demands can be served?” or “How must the grid be designed to serve the transmission demands of a specific production/consumption pattern?”.

\subsubsection*{2nd \jh{version:} 
adding rules for reducing production or consumption.}

Answering the questions mentioned above will not be very satisfying since production/consumption of nodes usually changes often over time and there may occur situations in which it can be necessary to reduce production/consumption of some nodes for some time to avoid line overload. Therefore, another \textit{research question} could be: “If a reduction is necessary, which reduction pattern should be applied?” For this question, the sustainant must be refined.

The new \textit{sustainant} could be the fulfillment of each consumer node’s demand. In order to call the overall system resilient, for each node, the delivered power needs to be within an acceptable range after the reduction.

To study the resilience of the power grid regarding this new sustainant, the current system model is insufficient. It has to be extended with the critical demand of each consumer node. The \textit{adverse influence} is then a continually changing production/consumption pattern on the nodes. 

Additionally, the system model is equipped with a first \textit{response option} to these pattern changes: an algorithm that specifies which nodes’ production or consumption gets reduced how and under which conditions. 

The new research interest can then be addressed by varying the reduction algorithm.

\subsubsection*{3rd \jh{version:} 
regarding multi-influences.}

In reality, of course, rules exist that deal with adverse influences which exceed the reaction capacity of the reduction algorithm or influence the sustainant in another way than only a changed infeed/consumption pattern. It may therefore be helpful to define an extended \textit{influence}, a multi-influence that consists of the well-known pattern changes as well as line tripping and generator failures. As a consequence, the system model has to be extended with the information whether a node or edge is active or not. A new interface to the environment is their activation/deactivation \cite{Plietzsch2015}. 

The new \textit{research question} could then be: Is the system’s reaction resilient (regarding the sustainant defined in the last version) in cases where the chosen reduction algorithm fails? To study this reaction, it is necessary to model the decisions taken in system operation (by engineers and software) in certain contingencies (\textit{response options}).

\subsubsection*{4th \jh{version:}
adding management options.}

If the research interest is not to determine whether the reaction of a specific system is resilient but rather which kind of management decisions make it resilient (exploration instead of prediction/evaluation), the model has to be extended by a set of additional \textit{response options} that can be chosen in certain contingencies. For example, the network operator could build additional lines, the government could introducing network fees, taxes or subsidies to incentivise changes in production or consumption, and electricity companies could change their pricing schemes and production locations.

This system model would not be purely deterministic or stochastic since the agents’ decisions are not modelled, only their decision options. Therefore, the model would have game-theoretical traits.

\subsubsection*{5th \jh{version:} 
frequency stability.}

In more recent considerations on the resilience of power grids, frequency stability has become an important aspect due to an increasing producer volatility caused by larger shares of renewable energy sources. Therefore, a suitable extension of the \textit{sustainant} is the following: At every node, the frequency must not leave an acceptable range for longer than some (very short) acceptable time spam so that devices do not get damaged \cite{Hellmann2016}. Adding this aspect to the already considered sustainant, we obtain a “multi-sustainant”. To answer any research question regarding this new sustainant, the electricity transport on the network has to be modelled representing dynamics of frequencies and phases on short time scales. Technically, this could be done by replacing power flow equations by so-called ``swing equations,'' the power flow would then be a result of these new equations.
The \textit{adverse influence} would then be extended by the change of frequency at a specific node due to fluctuations in consumption and renewable energy production,
and {\em response options} would include the programming of fast-reacting power electronics devices that switch lines and redirect flows.

\subsubsection*{Outlook on further modelling approaches}

The presented ways of how to answer the checklist when studying the resilience of a power grid are by far not complete. In order to get an idea of the vast number of other possibilities, consider these further aspects:
\begin{itemize}
    \item The net operator could define the purpose of the network (and, in their perspective, the sustainant) as generating profit. A corresponding model would have to include a power market which could produce internal fluctuations as an adverse influence inside of the system boundaries \cite{Heitzig2017}.
    \item 	From the perspective of the government of a country having a power grid, an interesting question could be: What policy-instruments give resilience-promoting incentives to grid-operators?  Such a question builds upon the answers of many of the other research questions (in order to know which management decisions of a grid operator would be resilience-promoting) but adds a model layer reflecting the mechanisms leading from policy-instruments to such decisions.
\end{itemize}
\small
\begin{table}[!ht]\small
    \centering
    \begin{tabular}{>{\raggedright}p{23mm}>{\raggedright}p{6mm}>{\raggedright}p{24mm}>{\raggedright}p{22mm}>{\raggedright}p{25mm}>{\raggedright\arraybackslash}p{31mm}}\toprule
         \bf Appli\-cation & \bf Ver\-sion & \bf System & \bf Sustainant & \bf Adverse influence & \bf Response options  \\\midrule
         Amazon rain\-forest & 1 & tree-covered area $C$ & $C>1/2$ & slowly rising aridity & convergence to stable equil.\ $C$ \\\addlinespace[0.3em]
         & 2 & & & + abrupt reduction in $C$ & \\\addlinespace[0.3em]
         & 3 & + species composition & & & + adaptation in species mix \\\addlinespace[0.3em]
         & 4 & + regional deforestation & & + econ. drivers of deforestation & + regulation against deforestation \\\midrule
         Fishery & 1 & fish stock, total harvest & total yield & population decline & convergence to stable equil.\ stock \\\addlinespace[0.3em]   
         & 2 & + total fishing effort & total yield & & + lowering of effort \\\addlinespace[0.3em]
         & 3 & + individual fishers & fishers' minimal yield & & + fishers' strategic interaction \\\addlinespace[0.3em]
         & 4 & + external market & income-based welfare metric & + price drop & + output reduction, taxation \\\midrule
         Electricity transmission system & 1 & network of transmission lines & every node's transmission demand is met & shifting consumption / production patterns & shifts in power flow due to Kirchhoff's laws\\\addlinespace[0.3em]
         & 2 & + production / consumption adjustment mechanisms & minimal level of consumption & & automatic production / consumption reduction\\\addlinespace[0.3em]
         & 3 & + possibility that lines or generators are down & & + line tripping, generator failure & + certain measures taken by system operators \\\addlinespace[0.3em]
         & 4 & + system operators & & & + pricing, regulation, building new lines \\\addlinespace[0.3em]
         & 5 & & + frequency stays within acceptable range & + fast fluctuations in renewable energy production & + power electronics \\
    \bottomrule
    \end{tabular}
    \caption{Possible answers to the guiding questions from our checklist for the three application examples. A ``+'' symbol implies  ``additional to the items directly above''.}
    \label{tab:other}
\end{table}

\end{document}